\documentclass[12pt]{JHEP3}

\usepackage{epsfig}

\def\spose#1{\hbox to 0pt{#1\hss}}
\def\ltapprox{\mathrel{\spose{\lower 3pt\hbox{$\mathchar"218$}}
 \raise 2.0pt\hbox{$\mathchar"13C$}}}
\def\gtapprox{\mathrel{\spose{\lower 3pt\hbox{$\mathchar"218$}}
 \raise 2.0pt\hbox{$\mathchar"13E$}}}

\title{Twisted Eguchi-Kawai Reduced Chiral Models}

\author{Stefano Profumo \\
        Scuola Internazionale Superiore di Studi Avanzati \\
	Via Beirut 2-4, I-34014 Trieste, Italy \\ 
	E-mail: \email{profumo@sissa.it} 
} 

\author{Ettore Vicari \\ 
	Dipartimento di Fisica dell'Universit\`a 
	di Pisa and I.N.F.N. \\
        Via Torricelli 2, I-56127 Pisa, Italy \\ 
	E-mail: \email{vicari@df.unipi.it} 
}

\abstract{
We study the twisted Eguchi-Kawai (TEK) reduction procedure for large $N$ unitary
matrix lattice models. In particular, we consider the case of two-dimensional principal
chiral models, and use numerical Monte Carlo (MC) simulations to
check the conjectured equivalence of TEK reduced model
and standard lattice model in the large-$N$ limit.
The MC results are compared with the
large-$N$ limit of lattice principal chiral models
to verify the supposed equivalence.
The consistency of the TEK  reduction procedure is verified
in the strong-coupling region, i.e. for $\beta<\beta_c$
where $\beta_c$ is the location of the large-$N$ phase transition.
On the other hand, in the weak-coupling regime
$\beta>\beta_c$, relevant for the continuum limit, 
our MC results do not support the equivalence of the large-$N$ limits of
the lattice chiral model and the corresponding TEK reduction. 
The implications for the correspondence between TEK model
and noncommutative field theory are also discussed. 
}

\keywords{Matrix Theories, 1/N Expansion, Lattice Quantum Field
Theories, Lattice Gauge Field Theories}

\begin{document}

\section{Introduction}

A renewed interest in reduced matrix models has been motivated by the
recent discovery of their relevance to noncommutative field
theories and to superstring matrix models, see, e.g.,
Refs.~\cite{BFSS-97,IKKT-97,NN-98,AIIKKT-00,IIKK-00,AMNS-00,AABHN-01,%
Makeenko-01,Szabo-01} and references therein.
Reduced matrix models were introduced for
the study of the large-$N$ limit
of $SU(N)$ gauge theories, see, e.g., Ref.~\cite{Das-87} for a review.
Eguchi and Kawai~\cite{EK-82} pointed out that, as a consequence of the 
large-$N$ factorization, one can construct one-point theories equivalent
to lattice $SU(N)$ gauge theories in the limit $N\rightarrow\infty$. 
The original proposal was just the replacement
of all link variables $U_\mu(x)$ of the Wilson formulation
of lattice gauge theories with four $SU(N)$ matrices 
according to the
reduction rule $U_\mu(x)\rightarrow U_\mu$. 
This simple reduction procedure was shown to fail
in the weak-coupling region,
because of the spontaneous breaking of a symmetry that is crucial 
for the equivalence \cite{BHN-82}. 
Modifications of the original Eguchi-Kawai 
model have been then proposed~\cite{BHN-82,EN-83,GO-83-2}.
The most promising one 
is represented by the so-called
twisted Eguchi-Kawai (TEK) reduction \cite{EN-83,GO-83},
which  replaces 
$U_\mu(x)\rightarrow D(x)U_\mu D(x)^\dagger$,
where $D(x)= \prod_\mu \Gamma_\mu^{x_\mu}$
and $\Gamma_\mu$ are appropriate $SU(N)$ matrices obeying the 't~Hooft
algebra.

Since the proposal of the application of the TEK reduction
prescription \cite{EN-83,GO-83}, the numerical checks 
\cite{GO-83-2,GK-83,FH-84,HM-86} of the
validity of the procedure have not been, to our opinion,
completely exhaustive. 
We believe that, also in view of the renewed interest for matrix
models, and particularly for TEK reduced versions,
it is important to perform
an accurate test of the conjectured equivalence
with the corresponding lattice matrix model.
This point may eventually 
be of relevance for the r\^{o}le of TEK models in
noncommutative field theories or in superstring
matrix (toy) models, where  the theory should be
analyzed in a limit different from the planar one.

In this paper we study the application of the
twisted Eguchi-Kawai (TEK) reduction procedure to 
two-dimensional principal chiral
models on the lattice \cite{EN-83,ABR-84,DK-84,GO-84}.
Two-dimensional chiral models and four-dimensional
gauge theories present interesting analogies, 
see, e.g., Refs.~\cite{GS-81,Polyakov-88,RCV-98,DPRV-02}.
For example, they
are both asymptotically free unitary matrix theories, and their
large-$N$ limit is represented by a formal sum of planar graphs.
In the large $N$ limit, one can apply 
the TEK reduction procedure to both lattice theories.
In both cases
the validity of the prescription is demonstrated 
by showing that the
Schwinger-Dyson equations for the reduced and the
non-reduced theory become formally the same in the limit $N=\infty$.
In two-dimensional
lattice chiral models a stringent test of the TEK reduction procedure 
should be simplified, for essentially three reasons:
a better knowledge of the large-$N$ limit of the lattice chiral model, 
an easier way to define observables in spin models,
and a simpler reduced model constituted by only one unitary matrix.

It has been recently realized that TEK reduced model may 
provide a nonperturbative formulation of noncommutative field theories
\cite{AIIKKT-00,IIKK-00,AMNS-00,Makeenko-01,Szabo-01}.
In particular, one may establish a correspondence 
between the TEK reduced model of the lattice principal chiral theory
and a noncommutative U(1) principal chiral model \cite{profumo}.
This is formally achieved in a continuum limit
that keeps the noncommutativity parameter 
$\theta_{\mu\nu}$ fixed \cite{AIIKKT-00,IIKK-00,AMNS-00,Makeenko-01,Szabo-01},
where $\theta_{\mu\nu}$ scales as $a^2 N$ and  $a$ is the lattice
spacing. Assuming
the equivalence of the TEK reduced and the commutative  principal
chiral model in the planar limit, 
one may rewrite this condition 
as a double scaling limit involving the lattice coupling and $N$ \cite{profumo}.
Therefore, our study of TEK reduced model in the planar limit
is also useful to clarify the correspondence
between TEK model and noncommutative field theory, and therefore
the possibility of achieving a nonperturbative lattice formulation
of a noncommutative field theory.

The plan of the paper is as follows. In Sec. \ref{Sez_SD} we
introduce the original field theory and its reduced counterpart,
and show on what grounds one expects that it coincides with the
non-reduced theory in the large $N$ limit. 
In Sec.~\ref{secnum} we present our numerical results
obtained by performing Monte Carlo simulations.
Finally, Sec.~\ref{Sez_conc} contains comments
and conclusions.


\section{TEK reduced chiral models}\label{Sez_SD}

\subsection{TEK reduced models}

Twisted Eguchi-Kawai reduced models are based on the 
idea that, when $N\to\infty$, the ${\rm SU}(N)$ group becomes so large
that it may accommodate the full Poincar\`e group as a subgroup, and 
allow us to find representations of the
translation and rotation operators among the elements of ${\rm
SU}(N)$.  As a consequence, one may reformulate the full
theory in terms of a finite number of matrix field variables defined
at a single space-time site (or on the $d$ links emerging from the
site in the case of a lattice gauge theory) and of the above-mentioned
representations of the translation group.  This reformulation is
called ``twisted Eguchi-Kawai'' reduced version of the theory
\cite{EN-83,GO-83}.

Consider a lattice field theory with fields $\phi(x)$ in the
adjoint representation of $SU(N)$. The twisted reduction
prescription reads:
\begin{equation}\label{prescription}
\phi(x) \ \rightarrow \ D(x) \ \phi \ D^\dag(x),
\end{equation}
where
\begin{equation}
D(x) \ = \ \prod_\mu(\Gamma_\mu)^{x_\mu},
\end{equation}
and $\Gamma_\mu$ are $SU(N)$ matrices obeying the so called t'Hooft algebra:
\begin{equation}
\Gamma_\mu\Gamma_\nu=Z_{\mu\nu}\Gamma_\nu\Gamma_\mu,
\end{equation}
where $Z_{\mu\nu}$ is an element of the center of the group,
$\mathbb{Z}_N$,
\begin{equation}\label{tensor}
Z_{\mu\nu}=\exp\left(\frac{2\pi i}{N}n_{\mu\nu}\right)
\end{equation}
and $n_{\mu\nu}$ is an integer valued antisymmetric $d\times d$
matrix (in $d$ dimensions). 
$\Gamma_\mu$ is the matrix implementing the translation by one lattice spacing
in the $\mu$ direction.
The reduced action is simply obtained
by substituing (\ref{prescription}) into the action of the
original lattice field theory:
\begin{equation}
S_{TEK}(\phi,n_{\mu\nu})=\frac{1}{\mathrm{vol}}S[D(x)\phi
D^\dag(x)],
\end{equation}
and the partition function reads:
\begin{equation}
Z_{TEK} \ = \ \int \mathcal{D}\phi \ \exp(-S_{TEK}),
\end{equation}
for a fixed value of $Z_{\mu\nu}$. The expectation value of
any functional of the reduced field $\phi$ is given by:
\begin{equation}
\langle O(\phi)\rangle_{TEK}=\frac{1}{Z_{TEK}}\int \mathcal{D}\phi
\ O(\phi)\exp(-S_{TEK})
\end{equation}
The correspondence between correlation functions of the reduced
model and of the original field theory is as follows. Let $f(x)$ be
any invariant functional of the field $\phi(x)$. Then
\begin{equation}\label{equivalence}
\langle f[\phi(x)] \rangle_{\mathrm{field \ theory}} \ = \ \langle
f[D(x)\phi D^\dag (x)]\rangle_{TEK}
\end{equation}

The actual check of validity
of the reduction procedure is based on the comparison
of the Schwinger-Dyson
equations of the reduced and original models.  
This procedure however
requires some attention, since the limit of infinitely many degrees of
freedom within the group itself allows the possibility of spontaneous
breakdown of some of the symmetries which would be preserved for any
finite value of $N$.  Moreover, large
$N$ is a thermodynamical limit: $N$ must go to infinity before any
other limit is considered, and sometimes the limiting procedures do
not commute.

\subsection{Principal chiral models}

The two-dimensional principal chiral field theory 
is defined through the following action
\begin{equation}
S={1\over T} \int d^2x \,{\rm Tr}\,\left[ \partial_\mu U(x) \partial_\mu
U^\dagger(x) \right],
\label{caction}
\end{equation}
where $U(x)$ are unitary matrix variables.
The action is naturally invariant under the transformations 
$U\rightarrow V_LU$ and $U\rightarrow UV_R$ where
$V_L,V_R \in U(N)$.
As amply discussed in the literature (see, 
e.g., Refs.~\cite{Polyakov-88,RCV-98,DPRV-02} and references therein),
two-dimensional chiral models and four-dimensional lattice gauge theories
manifest deep  analogies,  such asymptotic freedom, large-$N$ planar
limit, etc...

A corresponding  lattice principal chiral model can be obtained by the usual
substitution of the derivative with a finite difference,
obtaining
\begin{equation}\label{latticeaction}
S=-\beta N\sum_{x}\sum_{\mu=1,2}\mathrm{Tr} \
\Big[U_xU^\dag_{x+\mu}+U_{x+\mu}U^\dag_x\Big],
\end{equation}
where $\beta N \equiv 1/T$.
In view of a large-$N$ analysis one may consider both $SU(N)$ and $U(N)$
models, since they are expected to reproduce the same statistical
theory in the limit $N\rightarrow\infty$, which is meant at fixed $\beta$.
Moreover, $SU(N)$ and $U(N)$ chiral models have
the same continuum limit at any finite $N\geq 2$.

For a recent review of results 
concerning chiral models see, e.g., Ref.~\cite{RCV-98}.
Here we only mention 
the presence of a peculiar large-$N$ phase transition
\cite{GS-81-2,RV-94,CRV-95,CRV-95-2}
which manifests itself with a power-law singularity
in the $N\rightarrow\infty$ limit of the
specific heat,
i.e. $C_\infty \sim |\beta_c-\beta|^{-\alpha}$,
and with a peak in the specific-heat at finite $N$
that becomes sharper and sharper with increasing $N$
\cite{RV-94,CRV-95-2}.
The analysis of the $N=\infty$ strong-coupling
series \cite{CRV-95} provides the estimates
$\beta_c=0.3060(4)$ and $\alpha=0.27(3)$,
while the extrapolations to $N=\infty$
of the position of the specific-heat peak, 
determined by
Monte Carlo simulations for $SU(N)$ and $U(N)$,
gives $\beta_c=0.3957(3)$.

\subsection{The reduced version: symmetries and equivalence}
\label{symmeq}

Implementing (\ref{prescription}) into (\ref{latticeaction}) gives
the TEK reduced lattice chiral model action:
\begin{equation}\label{azioneTEK}
S_{TEK}=-\beta N\sum_{\mu=1,2}\mathrm{Tr}\Big[U\Gamma_\mu
U^\dag\Gamma^\dag_\mu+h.c.\Big].
\end{equation}
In the present case, $n_{\mu\nu}$ of equation (\ref{tensor}) is
generically, in $d=2$, of the form
\begin{equation}
n_{\mu\nu} \ \equiv \ \left(
\begin{array}{c c}
0 & M \\
-M & 0
\end{array}
\right) \ , \qquad M\in\mathbb{Z}.
\end{equation}
For a given $N$ and $M$ the solution to (\cite{thoofttwists}) is
provided, up to global $SU(N)$ transformations, by the $N\times N$
shift and clock matrices
\begin{equation}
\begin{array}{c}
S^{(M)}_{i,j} \ \equiv \ \delta_{i+M,j},\\
\\
C_{i,j} \ \equiv \ e^{\frac{2\pi i}{N}(i-1)}\delta_{i,j}.
\end{array}
\end{equation}
The two matrices $\Gamma_\mu$ are given respectively by $S$
and $C$.

While the original field theory is invariant under the
$SU(N)_L\times SU(N)_R$ group, the reduced theory is invariant
under the following two symmetries:
\begin{eqnarray}
&&U \ \rightarrow \ z\cdot U \qquad z\in \mathbb{Z}_N, \label{sym1}\\
&&U \ \rightarrow D(x) \ U \ D^\dag(x). \label{sym2}
\end{eqnarray}
The first symmetry corresponds to the contraction of
the original symmetry of the model to the center of the algebra of
the symmetry group, while the second one is reminiscent
of the translational invariance of the original theory.

In order to show the 
equivalence of the reduced theory with the
original field theory in the large $N$ limit,
the starting point is given by the 
Schwinger-Dyson equations, i.e. the equation obtained from a
generic correlation function in the original theory (the reduced
case is the obvious translation using Eq. (\ref{prescription}))
\begin{equation}
G^{(n)}=\langle\mathrm{Tr}\left(\prod_{i=1}^nU_{x_i}^{k_i}\right)\rangle,
\end{equation}
by the following steps:
\begin{enumerate}
\item substitute $U_{x_i}^{k_i}$ with $\lambda^aU_{x_i}^{k_i}$,
where $\lambda^a$ is a $SU(N)$ generator;
\item implement the change of variables $U\rightarrow
(1+i\alpha\lambda^b)U$ in the integral (implicitly taken in the
average), using the Haar measure invariance;
\item contract the $a,b$ indices and extract the linear part in
the parameter $\alpha$.
\end{enumerate}
By comparing the two sets of equations obtained in the original
and in the reduced theory, one finds 
manifestly corresponding terms and also some
extra terms in the reduced theory. These extra terms exactly
cancel if the symmetry 
$U \rightarrow D(a)UD^\dag(a)$ of the reduced model
is not spontaneously broken \cite{ABR-84}.
Therefore, the two models possess the same Schwinger-Dyson
equations if the above mentioned symmetry is not broken in any
regime, and in particular in the continuum limit. 
The point is then 
whether this
implies the equivalence  of the two models, i.e. that they have the same
expectation values of invariant quantities.
Since the large $N$ equivalence of the twisted reduced models is
based on the above arguments even for the lattice gauge theory case, it
is an important task to verify, if the invoked symmetry holds,
whether the Schwinger-Dyson equation based demonstration of the
equivalence is indeed conclusive and sufficient or not. 

The equivalence of the lattice chiral model and
the reduced TEK model in the $N=\infty$ limit 
is not actually questioned in the strong-coupling
region, i.e. for sufficiently small values of $\beta$.
In this region the symmetries of the TEK reduced model 
are unbroken and therefore the equivalence of
Schwinger-Dyson equations holds.
One may then argue that this is sufficient 
to yield the equivalence of the reduced and nonreduced
models.  In the strong-coupling domain, i.e.
within the convergence radius of the strong-coupling expansion,
expectation values are analytic in the coupling $\beta$, and 
their boundary
value at $\beta=0$ can be easily computed. 
As a consequence, it is formally possible to
solve the Schwinger-Dyson equations in terms of strong-coupling
series by iterative methods.
Since the $N=\infty$  boundary values at $\beta=0$ of 
the lattice chiral model and the reduced TEK model coincide,
equal Schwinger-Dyson equations imply the formal equivalence 
within the convergence radius of the strong coupling expansion
\footnote{Notice that the convergence radius of the strong coupling
expansion does not necessarily
coincide with the large-$N$ transition point $\beta_c$,
but it may be smaller.}.
Problems may occur in the weak-coupling side of
a large-$N$ phase transition.

It has been also argued that \cite{Das-87}
the reduced TEK model is, 
in a sense, equivalent to a theory on a box of size $L=N$.
In the large-$N$ limit the finite-$N$ corrections should be
$O(1/N^2)$, just as in the $SU(N)$ lattice chiral theory.
Since $N^2=L^2$, finite-$N$ corrections can be seen
as finite volume corrections. 
Therefore, one may expect that 
the asymptotic large-$N$ regime is reached
for $N>>\xi$, where $\xi$ is
the $N=\infty$ correlation length 
at the given value of $\beta$.
Note that in TEK reduced models
the large-$N$ and thermodynamic limits are 
connected and  approached simultaneously,
while in the standard lattice models 
one first performs the $L\rightarrow \infty$ limit
and then the large-$N$ limit. 
In a general field theory there are no a priori
reasons why these two limits should commute.

\section{Numerical results}
\label{secnum}

In order to study the large-$N$ limit of the TEK model, we
have performed Monte Carlo simulations 
for several values of $N$, up to $N=200$,
and for values of $\beta$ in the strong and weak coupling regime. 
We used a Metropolis algorithm to update the
$SU(N)$ matrix $U$. Trial matrices were selected by multiplying
the actual matrix $U$ by a random SU(2) matrix embedded in $SU(N)$,
choosing randomly among the $N(N-1)/2$ $SU(2)$ subgroups. 
More precisely, once the $SU(2)$ subgroup is randomly chosen, 
we performed ten Metropolis hits with an approximate
acceptance of 50\%.
Each $SU(2)$ updating requires $O(N)$ operations. 
The number of $SU(2)$-subgroup updatings per run was $O(10^9)$.
We should also mention that for large values of $N$,
$N\gtrsim 60$ say, we observed some problem of thermalization
when using completely random 
configurations (for example, constructed  by
multiplying $N(N-1)/2$ completely random SU(2) matrix  embedded in
a $N\times N$  and associated to different subgroups)
as starting point of our simulations.
This occurred
in both the strong- and weak-coupling regions,
worsening with increasing $\beta$.  So,
as starting point we used either moderately random matrices or
the unity matrix.
It is well known that a simple Metropolis algorithm
does not provide a particularly efficient method
to simulate a statistical system. Our choice was essentially due to
the fact that the reduced action is quadratic in the
matrix variable $U$. Moreover, it does not lend itself
to a linearization by introducing new matrix variables,
as in the case of the reduced TEK gauge theory \cite{FH-84}.

\TABLE[ht]{
\caption{Internal energy $E$ and magnetic susceptibility $\chi$
for $\beta=0.28$.}
\label{tab28}
\begin{tabular}{ccc}
\hline\hline
$N$ & $E$ &  $\chi$ \\
\hline
10   &   0.6155(2)  & 12.30(2)  \\
16   &   0.6346(1)  & 12.47(2)  \\
20   &   0.6409(1)  & 11.46(4)  \\
30   &   0.64766(7)  & 9.49(3)  \\
40   &   0.65046(8) & 8.44(3)  \\
50   &   0.65159(5) & 7.94(3)  \\
60   &   0.65226(4) & 7.65(3)   \\
80   &   0.65285(3) & 7.34(4)   \\
100  &   0.65329(4) & 7.21(4)   \\
150  &   0.65349(4) & 7.08(5)   \\
\hline\hline
\end{tabular}
}

In our simulations we measured the internal energy
\begin{equation}
E = 1 - \frac{1}{N} \langle {\rm Tr}\; 
U\Gamma_\mu U^\dagger \Gamma_\mu^\dagger \rangle,
\end{equation}
the quantity corresponding to the two-point function
$G(x) = \langle {\rm Tr}\; U(0) U(x)^\dagger \rangle$
using the equivalence (\ref{equivalence}),
and in particular its low-momentum components, such as
the (zero-momentum) magnetic susceptibility
\begin{equation}
\chi = \left| {\rm Tr}\, U \right|^2.
\label{chitek}
\end{equation}
In order to check for
possible spontaneous breaking of
the symmetries (\ref{sym1}) and (\ref{sym2}), 
we considered the two quantities
\begin{eqnarray}
&&S_1 = \frac{1}{N} \langle\mathrm{Tr} \, U \rangle, \label{s1} \\
&&S_2 = \frac{1}{N}  \langle \mathrm{Tr} \,
U\Gamma_1 U^\dag\Gamma_2^\dag \rangle.\label{s2}
\end{eqnarray}
Indeed, $S_1$ is not invariant under the transformation
(\ref{sym1}) (but it is invariant with respect to (\ref{sym2})),
while $S_2$ is not invariant under the transformation
(\ref{sym2}) (but it is invariant with respect to (\ref{sym1})).

Information on the large-$N$ limit of lattice chiral models
in the strong coupling region can be obtained from the 
$N=\infty$ strong-coupling expansion and 
extrapolation of Monte Carlo results for  the $SU(N)$ and $U(N)$  
unitary groups.
$N=\infty$ strong-coupling series are reported in
Refs.~\cite{GS-81,GS-81-2,RV-94,CRV-95}. For example,
the internal energy is known to 17th order and the magnetic
susceptibility to 15th order.
Monte Carlo simulations for 
relatively large values of $N$ (up to $N=30$)
can be found in Refs. \cite{RV-94,CRV-95-2}.
It is useful to consider both $SU(N)$ and $U(N)$ data
because their expectation values aproach the $N=\infty$ values
from opposite sides, and allow us to have a better control
of the large-$N$ limit.

As already mentioned, 
the equivalence of reduced and nonreduced models
is not questioned in the strong-coupling region.
However, Monte Carlo simulations in the strong coupling region
are still useful to study the convergence of the
large-$N$ limit of TEK, which is expected to
be controlled by powers of $N^{-2}$.

After verifying a general agreement between the 
large-$N$ TEK results and the $N=\infty$ strong-coupling
series for small $\beta$, we have performed 
longer runs for $\beta=0.28$ in order to obtain an accurate test.
This value of $\beta$ is not too far from the
large-$N$ transition point $\beta_c\approx 0.306$.
The large-$N$ limit of $E$ and $\chi$ 
of the lattice chiral model
can be estimated from the analysis of the
corresponding strong-coupling series \cite{CRV-95},
obtaining \footnote{
Following Ref.~\cite{CRV-95},
we used Dlog-Pad\'e approximants to resum the series of
$dE/d\beta\sim (\beta_c - \beta)^{-\alpha}$ and
$d {\rm ln} \chi / d\beta \sim (\beta_c - \beta)^{-\alpha}$ 
with $\beta_c\approx 0.306$ and $\alpha\approx 0.27$.
The internal energy and the magnetic suceptibility are then
obtained by integrating with respect to $\beta$.}
$E_\infty=0.65370(4)$ and $\chi_\infty=7.02(1)$.
Extrapolations
of the Monte Carlo results  reported in 
Refs.~\cite{RV-94,CRV-95-2}, using the fact that the 
corrections are expected to be powers of $N^{-2}$, 
provide consistent results, i.e.
\footnote{
For example, the simplest extrapolation using a straight line
$a + b N^{-2}$ passing by the data for the largest values of $N$
(i.e. $\chi(N=30)=7.032(5)$, $\chi(N=21)=7.069(5)$ for $SU(N)$ and
$\chi(N=21)=6.972(8)$, $\chi(N=15)=6.924(7)$  for $U(N)$)
gives $\chi(N) = 7.00(1) + 32(6) N^{-2}$ for $SU(N)$ 
and $\chi(N) = 7.02(2) - 22(5) N^{-2}$ for $U(N)$. 
\label{footnote}}
$\chi_\infty=7.00(2)$ for $SU(N)$,
$\chi_\infty=7.02(2)$ and $E_\infty=0.65373(4)$ for $U(N)$.
Moreover, we mention that the large-$N$ limit of the second-moment
correlation length is $\xi_\infty\approx 1.53$.
In Table \ref{tab28} we report $E$ and $\chi$ as obtained by our
simulations for several values of $N$, up to $N=150$.
In Figs.~\ref{b28ene} and \ref{b28chi} we plot them 
versus $1/N^{2}$, which is the order of the expected asymptotic
corrections.
There we also show the MC data obtained from simulations
of $SU(N)$ and $U(N)$ lattice chiral models. 
The figures show clearly
that the correct $N=\infty$ limit is approached.
To be more quantitative, fitting the data for $N\geq 30$
allowing for a $O(N^{-2})$ correction, we obtained
\begin{eqnarray}
&&E \approx  0.65377(3) - \frac{5.46(7)}{N^2} ,\\
&&\chi \approx  7.02(2) + \frac{2.24(4)\times 10^3}{N^2} ,
\end{eqnarray}
with acceptable $\chi^2/d.o.f\approx 1$.
This analysis shows that the correct large-$N$ limit is approached
with corrections of $O(N^{-2})$, as expected.
Note, however, that the $O(1/N^2)$ corrections are much larger than
those found for the lattice chiral models, by about a factor 
$10^2$ for $\chi$, see the footnote \ref{footnote}.

\FIGURE[ht]{
\epsfig{file=b28eneall.eps, width=12truecm} 
\caption{
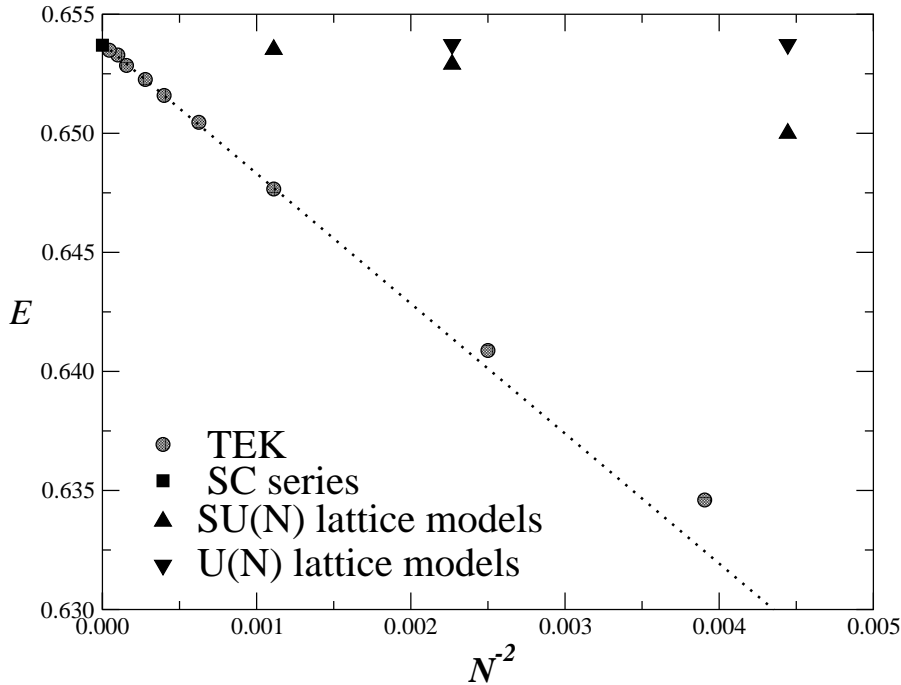
The internal energy versus $N^{-2}$ for $\beta=0.28$.
We also show the large-$N$ limit $E_\infty$ as obtained from the
analysis of the $N=\infty$ strong-coupling series, 
and  results from simulations of $SU(N)$ and $U(N)$
lattice chiral models. 
}
\label{b28ene}
}

\FIGURE[ht]{
\epsfig{file=b28chiall.eps, width=12truecm} 
\caption{
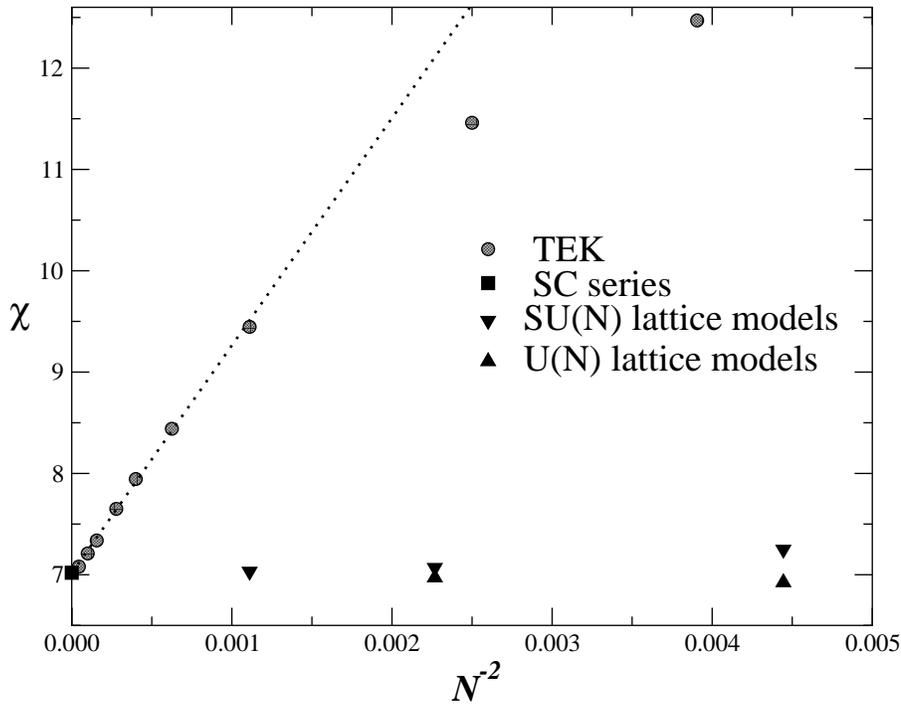
The magnetic susceptibility versus $N^{-2}$ for $\beta=0.28$.
We also show the large-$N$ limit $\chi_\infty$ as obtained from the
analysis of the $N=\infty$ strong-coupling series,
and  results from simulations of $SU(N)$ and $U(N)$
lattice chiral models.
}
\label{b28chi}
}

As already mentioned,
lattice unitary chiral models present a peculiar large-$N$ phase 
transition, which is characterized by a power-law singularity
of the specific heat at $N=\infty$ \cite{CRV-95}.
At finite $N$, numerical simulations show that the peak in the specific-heat
becomes sharper and sharper with increasing $N$ \cite{CRV-95-2}. 
A peak in the specific heat is also observed in the TEK reduced model:
it becomes sharper and sharper with increasing $N$. The
$N\rightarrow\infty$ extrapolation of its location provides 
the estimate $\beta_c\approx 0.306$,
in agreement with the results for the lattice chiral models.

These results provide an accurate check that 
for $\beta<\beta_c$ the TEK model shows the expected behavior,
i.e. it converges to the same large-$N$ limit of lattice chiral model, 
with corrections of $O(N^{-2})$.
We should also observe that, due to the large corrections,
the convergence to the $N=\infty$ limit
is rather slow, much slower than the convergence
of the lattice chiral models.
Comparing with the numerical effort
to reach a similar accuracy in simulations of lattice
chiral models, 
the reduced TEK model does not appear
to be convenient to numerically investigate the large-$N$ limit
of chiral models, at least in the strong-coupling domain.

\TABLE[ht]{
\caption{
We report the intenal energy $E$, the magnetic susceptibility $\chi$, and
the quantities $S_1$ and $S_2$, defined in Eqs.~(\protect\ref{s1})
and (\protect\ref{s2}), for $\beta=0.31, 0.32$ and several values of $N$.}
\label{tab31}
\begin{tabular}{cccccc}
\hline\hline
$\beta$ & $N$ & $E$ &  $\chi$ &  $S_1$ & Re $S_2$ \\
\hline
0.31 &   10   &   0.5251(1) & 22.06(2)  & $-0.0011(8) + i 0.0004(9)$ & 0.00002(3) \\
     &   16   &   0.5329(2) & 35.46(6)  & $-0.001(2) + i 0.001(2)$ & 0.00000(3) \\
     &   20   &   0.5342(2) & 43.9(2)  & $-0.004(3) + i 0.005(3)$ & $-$0.00001(5) \\
       & 30   &   0.5337(2) & 64.0(2)  & $0.003(2) - i 0.005(2)$ & $-$0.00002(3) \\
       & 40   &   0.5317(3) & 83.3(6)  & $0.001(4) + i 0.005(3)$ &  0.00006(4) \\
       & 60   &   0.5265(3) & 123(2)   & $-0.003(8) + i 0.003(6)$  &  0.00011(5) \\
       & 80   &   0.5236(2) & 157(2)   & $0.007(6) - i 0.001(5)$   & 0.00003(4) \\
       &100   &   0.5216(2) & 188(3)   & $-0.013(9) - i 0.014(8)$  &  0.00005(5) \\
       &150   &   0.5204(1) & 211(10)  & $-0.011(7) + i 0.004(10)$  &  0.00002(6) \\
       &200   &   0.5200(1) & 223(4)   & $-0.010(9) + i 0.004(9)$  &  0.00004(4) \\\hline
0.32 & 40     &   0.4902(1) & 148.2(7) & $-0.009(6) - i 0.002(5)$ & 0.00001(4) \\
     & 60     &   0.4880(2) & 235(4)   & $ 0.03(2) + i 0.03(2)$ & $-$0.00003(7) \\
     & 80     &   0.4869(2) & 331(3)   & $ -0.01(2) - i 0.03(2)$ & $-$0.00011(7) \\
     & 100    &   0.48656(4) & 404(2)   & $ 0.03(2) - i 0.02(2)$ & 0.00001(4) \\
     & 150    &   0.48629(8)& 565(7)   & $ 0.01(2) - i 0.06(2)$ & 0.00013(8) \\
\hline\hline
\end{tabular}
}

Let us now turn to the weak-coupling region, i.e.
for $\beta>\beta_c$. In this case information on the
large-$N$ limit of lattice chiral models 
comes only from the MC simulations of Refs.~\cite{RV-94,CRV-95-2}.
We performed runs for $\beta=0.31$ and $\beta=0.32$, for
which the $N=\infty$ correlation length are estimated
to be $\xi\approx 4$ and $\xi\approx 6$, respectively.
MC simulations up to $N=200$ for these values of $\beta$
should provide a convincing check of the large-$N$ convergence,
which should be  controlled by the ratio $N/\xi$
according to the arguments mentioned in Sec.~\ref{symmeq}.
The results of the simulations for $E$, $\chi$, and  the two
quantities $S_1$ and $S_2$, cf. Eqs.~(\ref{s1}) and (\ref{s2}),
are reported in Table \ref{tab31}.
They should be compared with the
the extrapolations to $N\rightarrow \infty$ of 
the MC results reported in 
Ref. \cite{RV-94,CRV-95-2}, that are
$\chi_\infty=34.1(2)$ and $E_\infty\approx 0.519$ for $\beta=0.31$,
and $\chi_\infty\approx 65$ and $E_\infty\approx 0.485$ for $\beta=0.32$.
Our MC data  are also plotted in
Figs.~\ref{b31ene} and \ref{b31chi}, where we also show the
MC results for $SU(N)$ and $U(N)$ lattice chiral models, taken
from Refs.~\cite{RV-94,CRV-95-2}.
While the data for the energy seem somehow
to approach the corresponding  $E_\infty$
(although it is not evident that the corrections are $O(N^{-2})$), 
those for  $\chi$ do not show any evidence
of convergence to $\chi_\infty$.
$\chi$ appears to increase with $N$ up to $N=200$.
Similar results are obtained for $\beta=0.32$.
We have also verified that the reduced
two-point function does not  exponentially
decrease at large distance, explaining
the behavior of its space integral $\chi$.
Finally, note that the results for $S_1$ and $S_2$, see Table~\ref{tab31}, 
are always
consistent with zero, so that they do not give any
indication of breaking of the symmetries (\ref{sym1})
and (\ref{sym2}).

\FIGURE[ht]{
\epsfig{file=b31eneall.eps, width=12truecm} 
\caption{
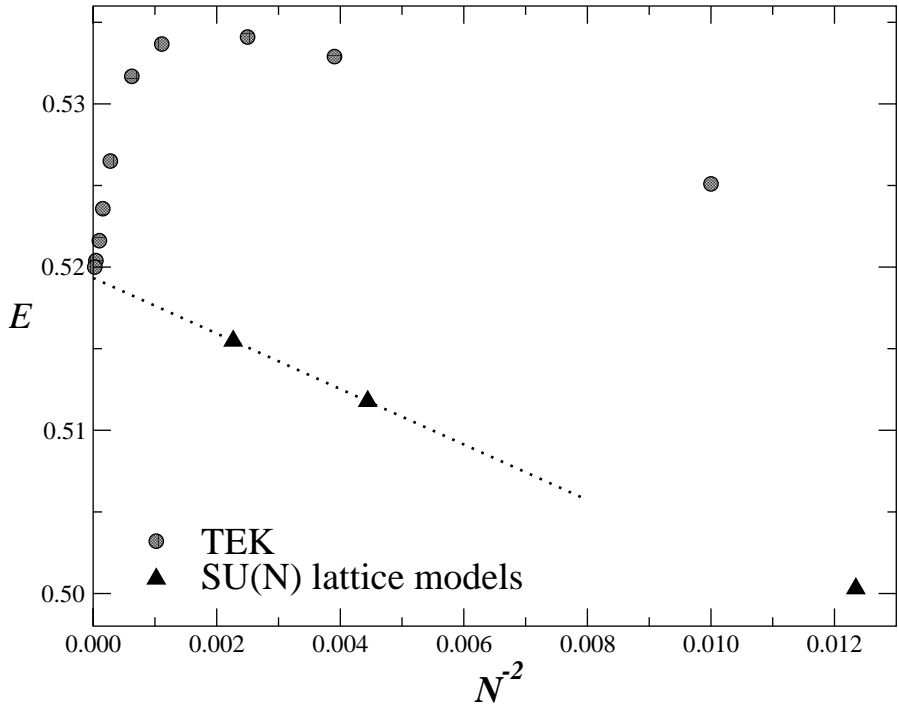
The internal energy versus $N^{-2}$ for $\beta=0.31$.
We also show the MC data for $SU(N)$ lattice chiral models,
and their large-$N$ extrapolation.
}
\label{b31ene}
}

\FIGURE[ht]{
\epsfig{file=b31chiall.eps, width=12truecm} 
\caption{
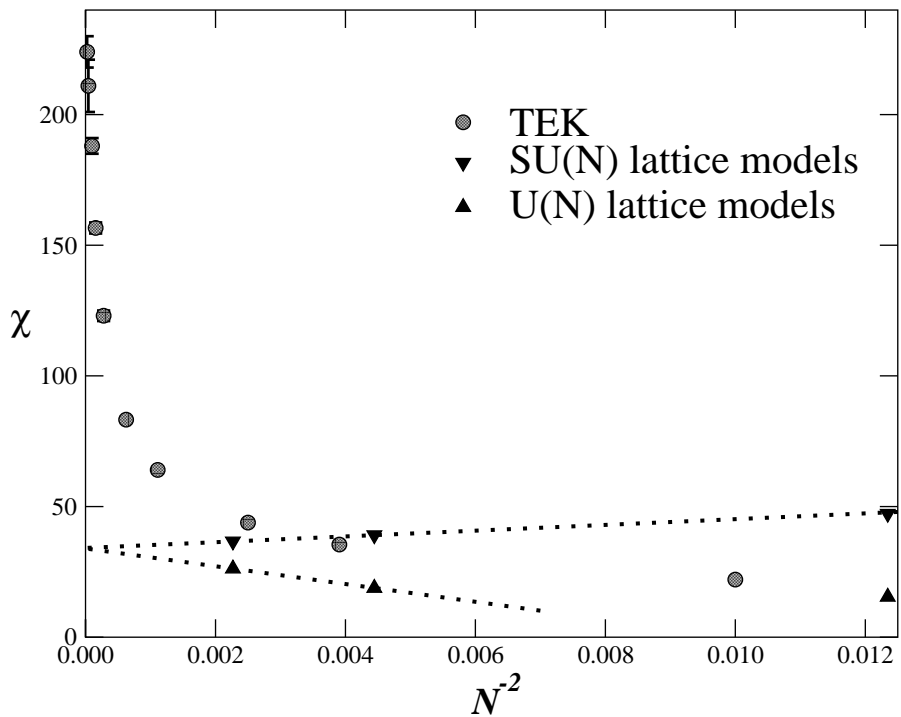
The reduced magnetic susceptibility for $\beta=0.31$ versus $N^{-2}$.
We also show the MC data obtained by simulating 
$SU(N)$ and $U(N)$ lattice chiral models; the lines 
represent their large-$N$ extrapolations.
}
\label{b31chi}
}

Our results apparently contradict earlier numerical works
\cite{DK-84,GO-84}, which were based on a much smaller statistic,
and where the comparison was essentially
restricted to the internal energy and the quantity $S_1$.

\section{Conclusions}\label{Sez_conc}

We have investigated  the application of the
twisted Eguchi-Kawai reduction procedure to 
two-dimensional principal chiral
models on the lattice, as general testcase
of the TEK reduction procedure.

In the strong coupling region, i.e. for $\beta<\beta_c$,
the large-$N$ limit of the reduced
theory appears to reproduce the results
available for the original field theory in the large-$N$ limit, 
and that  the TEK reduced model approaches the large-$N$ limit
with power corrections of $N^{-2}$.
On the other hand, 
the validity of the reduction prescription apparently
breaks down for $\beta>\beta_c$. 
In particular, it is not supported by the MC results for 
the magnetic susceptibility, that do not show evidence 
of convergence up to  $N=200$ and for $\beta$ values corresponding
to large-$N$ correlation lengths $\xi\approx 4$ and $\xi\approx 6$
(estimated extrapolating MC data for lattice chiral models).
Of course,
we cannot exclude that the correct asymptotic regime
is eventually set for larger values of $N$, $N\gg 10^2$ say. 
Although we note that
already for smaller values of $\beta\lesssim \beta_c$
the approach to the large-$N$ limit is rather slow,
in order to explain our results
we should suppose a strong slowing down of the convergence
with increasing $\beta$. 
In this regard, we should also add that
the equivalence of the theories in the
continuum limit requires a uniform convergence 
to $N=\infty$ when approaching the continuum limit.
Another possibility 
is that our numerical simulations have been
trapped by nontrivial extrema of the action, 
which has nothing to do
with the correct limit related to the chiral field theory. 
However, our numerical checks 
definitely exclude it at least for $N\lesssim 60$.

We have also checked for possible 
spontaneous breakings of the symmetries 
(\ref{sym1}) and (\ref{sym2}) of the TEK reduced model,
by computing the quantities $S_1$ and $S_2$, cf. Eqs.~(\ref{s1})
and (\ref{s2}), which are respectively
not invariant with respect to the symmetry (\ref{sym1}) and (\ref{sym2}).
Their expectation values appear to vanish, thus
suggesting that the symmetries of TEK are not
spontaneously broken, although one cannot exclude breaking of
symmetries associated with more complicated order parameters.
We recall that one of such symmetries, i.e. the symmetry (\ref{sym2}),
is essential for the demostration of the equivalence
of the $N=\infty$ Schwinger-Dyson equations.
Assuming that the Schwinger-Dyson equations become
formally equal for $N=\infty$, the question to be addressed is whether
this is sufficient to identify a theory.
Our numerical results  suggest that this
could not be the case, under the hypothesis that one can recover
the $N=\infty$ theory from the
finite $N$ behaviour of the reduced theory in its
$N\rightarrow\infty$ limit. 
In other words, the $N=\infty$ TEK reduced theory
may be still equivalent to the large-$N$ limit of lattice models,
but the limit is singular, so that it  cannot be
reached from finite values of $N$.
This issue is of relevance in any
application of the TEK reduction procedure, as in the case of
the lattice gauge theory.

The apparent failure of the TEK reduced theory 
to reproduce the planar limit of lattice chiral models 
does not necessarily prejudice the possibility
of considering the TEK theory system in a different limit,
possibly relevant to the fields where twisted Eguchi-Kawai reduced
models were recently applied: noncommutative field theories and
superstring matrix models. 
Indeed, TEK reduced chiral models with symmetry group $U(N)$ are
formally equivalent to a U(1)
noncommutative theory \cite{profumo}. 
This equivalence requires a nontrivial continuum limit that is
different from the planar
limit we have considered in this work, since one needs to keep the 
noncommutativity parameter $\theta_{\mu\nu}$ fixed, and
$\theta_{\mu\nu}$ scales as $a^2 N$ where $a$ is the lattice spacing.
Thus, the relevant continuum limit should be realized
for $a\rightarrow 0$ and $N\rightarrow \infty$ keeping $a^2 N$ fixed.
The lattice spacing is somehow controlled by the coupling $\beta$,
which should be appropriately tuned to realize the limit $a\rightarrow
0$. The relation between $a$ and $\beta$ is therefore essential
to define the continuum limit associated with the noncommutative
field theory. 
However, the apparent failure of TEK model
to reproduce the planar limit of the principal chiral theory
does not allow us to use the known asymptotic behavior  of the commutative
theory in the continuum limit, for which asymptotic freedom and
renormalization group imply $a\sim \exp (-8\pi\beta)$
in the continuum limit $a\rightarrow 0$ and for $N=\infty$.
Of course, this point deserves further investigation.

\acknowledgments{
We thank Luigi Del Debbio, Andrea Pelissetto, and
Paolo Rossi for useful discussions.}


\newpage


\begin{thebibliography}{99}


\bibitem{BFSS-97}
T. Banks, W. Fischler, S. H. Shenker, and L. Susskind,
\prd{55}{1997}{5112}.

\bibitem{IKKT-97}
N. Ishibashi, H. Kawai, Y. Kitazawa and A. Tsuchiya, 
\npb{498}{1997}{467}.



\bibitem{NN-98}
Nakajiama and J. Nishimura, 
\npb{528}{1998}{355}.

\bibitem{AIIKKT-00}
H. Aoki, N. Ishibashi, S. Iso, H. Kawai,
Y. Kitazawa and T. Tada, \npb{565}{2000}{176}.

\bibitem{IIKK-00}
H. Aoki, S. Iso, H. Kawai, and Y. Kitazawa,
\npb{573}{2000}{573}.

\bibitem{AMNS-00} 
J. Ambj\o rn, Y.M. Makeenko, J. Nishimura and R.J. Szabo, 
\jhep{11}{1999}{029};
\plb{480}{2000}{399};
\jhep{05}{2000}{023};
\jhep{07}{2000}{013}.


\bibitem{AABHN-01}
J. Ambj\o rn, K.N. Anagnostopoulos, W. Bietenholz, T. Hotta, and
J. Nishimura, \jhep{07}{2000}{011}.

\bibitem{Makeenko-01}
Y. M. Makeenko, \jetpl{72}{2000}{393}.

\bibitem{Szabo-01} 
R. J. Szabo, \emph{Quantum field theory on noncommutative spaces}, 
\hepth{0109162}.

\bibitem{Das-87}
S. R. Das, \rmp{59}{1987}{235}.

\bibitem{EK-82}
T. Eguchi and H. Kawai, \prl{48}{1982}{1063}.

\bibitem{BHN-82}
G. Bhanot, U. M. Heller, and H. Neuberger, \plb{115}{1982}{237}.

\bibitem{EN-83}
T. Eguchi and R. Nakayama, \plb{122}{1983}{59}.

\bibitem{GO-83-2} A. Gonzales-Arroyo and M. Okawa,
\plb{133}{1983}{415}. 

\bibitem{GO-83} A. Gonzales-Arroyo and M. Okawa, 
\prd{27}{1983}{2397}.

\bibitem{GK-83} A. Gonzales-Arroyo and C. P. Korthals Altes,
\plb{131}{1983}{336}.

\bibitem{FH-84}
K. Fabricius and O. Haan, \plb{143}{1984}{459}.

\bibitem{HM-86}
O. Haan and K. Meier,
 \plb{175}{1986}{192}.


\bibitem{ABR-84} B. Aneva, Y. Brihaye and P. Rossi, \plb{134}{1984}{245}.

\bibitem{DK-84} S. R. Das and J. B. Kogut, \npb{235}{1984}{521}.

\bibitem{GO-84} A. Gonzales-Arroyo and M. Okawa, \npb{247}{1984}{104}.

\bibitem{GS-81} 
F. Green and S. Samuel, \npb{190}{1981}{113}.

\bibitem{Polyakov-88}
A. M. Polyakov, {\em Gauge Fields and Strings},
Harwood Academic Publishers, New York 1988.

\bibitem{RCV-98}
P. Rossi, M. Campostrini, and E. Vicari,
\prep{302}{1998}{143}.

\bibitem{DPRV-02}
L. Del Debbio, H. Panagopoulos, P. Rossi, and E. Vicari,
\prd{65}{2002}{021501}; \jhep{01}{2002}{009}.

\bibitem{profumo} S. Profumo, 
\emph{Noncommutative Principal Chiral Models}, 
\hepth{0111285}.

\bibitem{GS-81-2} 
F. Green and S. Samuel, 
\plb{103}{1984}{110}.

\bibitem{RV-94} 
P. Rossi and E. Vicari, 
\prd{49}{1994}{1621}; \prd{49}{1994}{6072};
\prd{55}{1997}{1698} (E).


\bibitem{CRV-95}
M. Campostrini, P. Rossi and E. Vicari, 
\prd{51}{1995}{958};
\prd{52}{1995}{358};
\prd{52}{1995}{386}.

\bibitem{CRV-95-2}
M. Campostrini, P. Rossi and E. Vicari, 
\prd{52}{1995}{395}.

\bibitem{thoofttwists} G.'t Hooft, 
\cmp{81}{1981}{529}.


\end{thebibliography}
\end{document}